\documentclass[12pt,epsf]{article}

\usepackage[dvips]{graphicx}

\newcommand{\beq}{\begin{equation}}
\newcommand{\eeq}{\end{equation}}
\newcommand{\bea}{\begin{eqnarray}}
\newcommand{\eea}{\end{eqnarray}}
\begin{document}
\thispagestyle{empty}
\vspace*{-15mm}
{\bf OCHA-PP-326}\\

\vspace{15mm}
\begin{center}
{\LARGE\bf
Is it possible to unify three kinds of Dark Matters into a Kaluza-Klein Neutrino? 
}
\vspace{7mm}

\baselineskip 18pt
{\bf Sakiko G. J. Nishio$^{1}$, Kazuharu Bamba$^{1, 2}$ and Akio Sugamoto$^{1}$}
\vspace{2mm}

{\it
$^1$Department of Physics, Graduate School of Humanities and Sciences, \\
   Ochanomizu University, Tokyo 112-8610, Japan}\\
{\it
$^2$Leading Graduate School Promotion Center, \\
Ochanomizu University, Tokyo 112-8610, Japan}\\
\vspace{10mm}
\end{center}
\begin{center}
\begin{minipage}{14cm}
\baselineskip 16pt
\noindent
\begin{abstract}
A unified theory of including all kinds of dark matters into a single species (field) is discussed. In particular, it is considered that the Warm Dark matter (WDM), the existence of which may be required by 
the detailed $N$-body simulations of galaxies using the $\Lambda$CDM model, 
is the right-handed neutrino, the Hot Dark Matter (HDM) is 
the left-handed neutrino, and the Cold Dark Matter (CDM) is 
the first Kaluza-Klein (KK) mode of neutrino. 
The study on how to detect the first KK neutrino mode as CDM by LHC, SuperK, and IceCube is also explained.  Not only these detectors but also the recent experiments, such as DAMA/LIBRA, PAMELA, XENON100, and XMASS as well as the various satellite detectors including Planck should be examined. 
\end{abstract}


\end{minipage}
\end{center}

\baselineskip 18pt
\def\thefootnote{\fnsymbol{footnote}}
\setcounter{footnote}{0}

\vspace{15mm}

\section{Introduction}
Owing to the observational data taken by the COBE \cite{COBE}, WMAP \cite{WMAP}, and Planck~\cite{Ade:2013lta, Ade:2013uln} satellites, the nature of ingredients of our universe become more and more clear recently. Suppose that the universe is flat, it is considered that there exist dark energy, dark matter, and baryons in the universe (for a recent review on cosmology, see, e.g.,~\cite{Bamba:2012cp}). 

Regarding dark matter, it is known that in the $\Lambda$CDM model, the consequences obtained from the $N$-body simulations of galaxies imply the existence of Warm Dark matters (WDM)~\cite{WDM}\footnote{There have been reported a number of studies on the cosmological investigations on WDM in the literature, for instance, more recent references~\cite{Recent warm dark matter}.}. Moreover, the most plausible candidates for the Hot Dark Matter (HDM) are the left-handed neutrinos. Accordingly, suppose that the right-handed sterile neutrino is the WDM, we are urged to let the first Kaluza-Klein (KK) mode of neutrino be the Cold Dark Matter (CDM) in the KK higher-dimensional theory~\cite{K-K}. Then, there appears a unified scenario for three kinds of dark matters into a single species (or a single field). Various observational and theoretical aspects on dark matter have been reviewed in Refs.~\cite{Rev-DM}. 
As another candidate for dark matter in the framework 
of a higher-dimensional space-time theory, in Refs.~\cite{Prof-Hosotani-DM} 
there has been proposed dark matter in the so-called gauge-Higgs unification 
scenario. 

In this paper, we investigate a unification of all kinds of dark matters, namely, HDM, WDM, and CDM, in the KK five-dimensional space-time theory at a primitive level. In addition, we explore the way of detection of the first KK neutrino mode, corresponding to CDM, by using the detectors like Large Hadron Collider (LHC)~\cite{LHC}, Super-Kamiokande (SK)~\cite{Young:1995iv}, and IceCube~\cite{IceCube}. 
The units of $k_\mathrm{B} = c = \hbar = 1$ are used 
and the gravitational constant is expressed by $G_\mathrm{N}$  
with the Planck mass of $M_{\mathrm{Pl}} = G_\mathrm{N}^{-1/2} = 1.2 \times 
10^{19}$\,\,GeV. 

The paper is organized as follows. 
In Sec.~2, we explain Hot, Warm and Cold Dark Matters. 
In Sec.~3, we describe our scenario of unification of three kinds of 
dark matters and examine the direct detectability of the dark matters. 
Conclusions are described in Sec.~4. 
In Appendix A, we also give the formulae of the cross sections in terms of dark matter detection at LHC, SK, and IceCube.

\section{Hot, Warm and Cold Dark Matters}

We assume the flat Friedmann-Lema\^{i}tre-Robertson-Walker (FLRW) space-time with the metric 
%
$
ds^2 = -dt^2 + a^2(t) \sum_{i=1,2,3}\left(dx^i\right)^2 
$ with the scale factor $a (t)$. 
In this background, 
the following Freedman equation determines how the sphere of radius $a$ in the universe expands~\cite{Kolb and Turner};
\beq
H^2=\left(\frac{\dot{a}}{a}\right)^2=\frac{8\pi G_\mathrm{N}\rho}{3}-\frac{k}{a^2}+\frac{\Lambda}{3},
\eeq
where $\rho$ is the energy density, 
$k$ is the cosmic curvature and and $\Lambda$ is the cosmological constant. 
This equation gives the quantity of the ingredients, namely
\beq
1=\sum_{i}\Omega_{i}+\Omega_{k}+\Omega_{\Lambda} \,,
\eeq
where $\Omega_{i}=8\pi G_\mathrm{N}\rho_{i}/3H^2=\rho_{i}/\rho_c$ is called the density parameter of the ingredient \{$i$\} which denotes the ratio of energy density $\rho_i$ of the ingredient by the critical density $\rho_\mathrm{c}$, while $\Omega_{k}$ and $\Omega_{\Lambda}$ are the density parameters of the curvature [$k=0$ (flat), $1$ (closed), $-1$ (open)] and of the cosmological constant (``dark energy (DE)"), respectively. The critical density has an ambiguity $h^2$ coming from the Hubble constant $H^2$, so that the observed data (68 $\%$ CL) are summarized like \cite{Ade:2013lta, PDG2011} 
\bea
\Omega_\mathrm{total~matter} h^2&=&0.1199 \pm 0.0027 \,, \\ 
\Omega_\mathrm{baryon} h^2&=&0.02205 \pm 0.00028 \,, \\
\Omega_\mathrm{neutrino} h^2&=& \frac{\sum m_{\nu}}{93 \mathrm{eV}} \approx 0.001 \,, 
%
%
\eea
and for Dark Energy
\beq
\Omega_{\Lambda} = 0.686 \pm 0.020 \,,  
\eeq
where the Hubble constant is given by $H=100h$ km/s/Mpc with the ambiguity factor $h=0.673 \pm 0.012$, and $m_{\nu}$ is the mass of neutrinos. 
(It is seen that $\Omega$ is about $2\times \Omega h^2$.) 

The difference between $\Omega_\mathrm{total~matter}$ and $\Omega_\mathrm{baryon}$ may be attributed to the dark matters.  Then, provided that the current universe is flat, the density parameter of the total dark matters in total is
\bea
\Omega_\mathrm{DM}h^2=0.09785\pm 0.00242 \,.
\eea
Neutrino is a ``Hot Dark Matter (HDM)", 
smoothing the large scale structure of the universe. 
Now the structure of the universe is well understood by the large values of $\Omega_{\Lambda}$ and $\Omega_\mathrm{CDM}$, called the $\Lambda$CDM model. 
However, when the detailed structure of galaxies is examined, the ``Warm Dark Matter (WDM)" may be necessary \cite{WDM}.  It is because the simulated inner mass density profile of the DM halos by the $\Lambda$CDM model is more cuspy than the inferred profile by the rotation curve and others, and the satellite galaxies in the halo obtained by this $\Lambda$CDM simulation are more abundant than the observation in the Local Group \cite{Viel}.  In order to moderate the inside structure of galaxies there should be particles whose free streaming length is order of Mpc (size of galaxies), with a mass of order 1 keV.  This is the necessary reason of the WDM, the candidates of which are the right-handed (sterile) neutrino or gravitino \cite{Silk, Peebles, Pagels}.

\section{Unification of three kinds of Dark Matters}

To unify the Hot, Warm and Cold Dark Matters and put them into a neutrino species, we adopt the Kaluza-Klein (KK) theory~\cite{K-K, KK-theory} of the universal extra dimensions \cite{KK}, and attribute the zeroth KK modes of neutrinos to HDM ($\nu_L$) and WDM ($\nu_R$), and the first KK modes to CDM.

\subsection{Unified scenario}

A five dimensional free Dirac action is 
\bea
&& 
S=\int~d^3x \int dy~ \left[ \bar{\psi}(x, y) \left( i \gamma^{\mu} \partial_{\mu} +\gamma^{5} \frac{\partial}{\partial y}-\phi (y) \right) \psi (x, y)
\right. 
\nonumber \\
&& \left. 
{}-{\psi}(x, y)^{T}C \gamma^{5} \Phi (y) \psi (x, y)\right] \,,
\eea
where the last term is the Majorana mass term with the charge conjugation matrix being $C(=i \gamma^{0} \gamma^{2})\times \gamma^5$ \cite{Sato}, and 
we have introduced the $y$-dependent Higgs fields or the expectation values.  Here the fifth dimension is compactified to $S^1/Z_2$, that is, the circle with radius $R$ parametrized by $y= 0-2\pi R$ is folded to $y= 0-\pi R$ by a $Z_2$ projection $P: y \rightarrow -y$.  This $Z_2$ projection is the parity operation for the fifth dimension, and so $P: \psi(x, y) \rightarrow \gamma_5 \psi(x, -y)$ for the fermionic field. 

The fields can be expanded into the KK modes,
\bea
\psi(x, y)=\frac{1}{(\pi R)^{1/2}} \left[ \psi^{(0)}(x)+ \sqrt{2} \sum_n \left( \psi_1^{(n)}(x) \cos \left(\frac{n}{R} y\right) + \psi_2^{(n)}(x) \sin \left(\frac{n}{R} y\right) \right) \right] \,, 
\eea 
\bea
\phi(y)=\frac{1}{(\pi R)^{1/2}} \left[ \phi^{(0)}+ \sqrt{2} \sum_n \left(\phi_1^{(n)}\cos \left(\frac{n}{R} y\right) + \phi_2^{(n)} \sin \left(\frac{n}{R} y\right) \right) \right] \,, 
\eea
and the same expansion for $\Phi(y)$. 
Here, the superscript $(0)$ means the quantities appearing in the ordinary 
four-dimensional theory and the superscript $(n)$ shows the quantities of 
the KK particles of the $n$-th mode. 
If we classify the fields according to $P=\pm$, $\psi^{\pm}$, we have
\bea
\psi_{+}(x, y)&=&\frac{1}{(\pi R)^{1/2}} \left[ \psi^{(0)}_{R}(x)
\right. 
\nonumber \\
&& \left. 
{}+ \sqrt{2} \sum_n \left( \psi_{1R}^{(n)}(x) \cos \left(\frac{n}{R} y\right) + \psi_{2L}^{(n)}(x) \sin \left(\frac{n}{R} y\right) \right) \right] \,, \\
\psi_{-}(x, y)&=&\frac{1}{(\pi R)^{1/2}} \left[ \psi^{(0)}_{L}(x)
\right. 
\nonumber \\
&& \left. 
{}+ \sqrt{2} \sum_n \left( \psi_{1L}^{(n)}(x) \cos \left(\frac{n}{R} y\right) + \psi_{2R}^{(n)}(x) \sin \left(\frac{n}{R} y\right) \right) \right] \,,
\eea
and 
\bea
\phi_{+}(y)&=&\frac{1}{(\pi R)^{1/2}} \left[ \phi^{(0)}+ \sqrt{2} \sum_n \left(\phi_1^{(n)}\cos \left(\frac{n}{R} y\right) \right) \right] \,, \\
\phi_{-}(y)&=&\frac{1}{(\pi R)^{1/2}} \left[ \sqrt{2} \sum_n  \phi_2^{(n)} \sin \left(\frac{n}{R} y\right) \right] \,.
\eea

  Then, mass terms of the fermion are obtained as follows:
\bea
\hspace{-5mm}
{\cal L}_\mathrm{m}&=&\int^{\pi R}_{0} dy~\left[\bar{\psi}(x, y) \left(\gamma^{5} \frac{\partial}{\partial y}+\phi (y) \right) \psi (x, y)+{\psi}(x, y)^{T}C \gamma^{5} \Phi (y) \psi (x, y)\right] \\
\hspace{-5mm}
&=&\sum_{n \ge 1} \left(\frac{n}{R} \right)(\bar{\psi}^{(n)}_{1R} \psi^{(n)}_{2L}+\bar{\psi}^{(n)}_{2R} \psi^{(n)}_{1L})+ (\bar{\psi}^{(0)}_{R} \psi^{(0)}_{L})\frac{2}{\pi R}\int dy \phi_{-}(y) + (\mathrm{h.\,c.})\nonumber\\
\hspace{-5mm}
&+&\left(\psi^{(0)T}_{R} C\psi^{(0)}_{R}-\psi^{(0)T}_{L} C\psi^{(0)}_{L} \right)\frac{2}{\pi R}\int dy \Phi_{+}(y) \nonumber \\
\hspace{-5mm}
&+&
\sum_{n \ge 1}\left(\bar{\psi}^{(0)}_{R} \psi^{(n)}_{1L} \right) \frac{2}{\pi R}\int dy\cos \left(\frac{n}{R} y\right) \phi_{-}(y) \nonumber \\
\hspace{-5mm}
&+& \sum_{n \ge 1} \left( \bar{\psi}^{(0)}_{R} \psi^{(n)}_{2L} \right) \frac{2}{\pi R}\int dy\sin \left(\frac{n}{R} y\right) \phi_{-}(y) \nonumber \\
\hspace{-5mm}
&+& \sum_{m, n \ge 1}\left(\bar{\psi}^{(m)}_{1R} \psi^{(n)}_{1L} \right) \frac{2}{\pi R}\int dy\cos \left(\frac{m}{R} y\right)\cos \left(\frac{n}{R} y\right) \phi_{-}(y) 
\nonumber \\
\hspace{-5mm}
&+& \sum_{m, n \ge 1} \left( \bar{\psi}^{(m)}_{2R} \psi^{(n)}_{2L} \right) \frac{2}{\pi R}\int dy\sin \left(\frac{m}{R} y\right)\sin \left(\frac{n}{R} y\right) \phi_{-}(y) + (\mathrm{h.\,c.}) \nonumber \\
\hspace{-5mm}
&+&\left( \psi^{(0)T}_{R} C\psi^{(n)}_{1R} - \psi^{(0)T}_{L} C\psi^{(n)}_{1L} \right) \frac{2}{\pi R}\int dy \cos \left(\frac{n}{R} y\right) \Phi_{+}(y) \nonumber \\
\hspace{-5mm}
&+& \sum_{m, n \ge 1}\left(\psi^{(m)T}_{1R} C\psi^{(n)}_{1R} -\psi^{(m)T}_{1L} \psi^{(n)}_{1L}\right) \frac{2}{\pi R}\int dy \cos \left(\frac{m}{R} y\right)\cos \left(\frac{n}{R} y\right) \Phi_{+}(y)  \nonumber \\
\hspace{-5mm}
&+& \sum_{m, n \ge 1} \left( -\psi^{(m)T}_{2L} C\psi^{(n)}_{2L}+\psi^{(m)T}_{2R} \psi^{(n)}_{2R} \right) \frac{2}{\pi R}\int \sin \left(\frac{m}{R} y\right)\sin \left(\frac{n}{R} y\right) \Phi_{+}(y) \,. 
\eea
The first three terms in (16) give the KK masses for the $n$-th KK modes, the Dirac mass and the Majorana mass for the zero mode.  However, as we have assumed above, if the scalar fields $\phi(y)$ and  $\Phi(y)$ may depend on $y$ classically, that is, if the brane located at $y=0$ may be solitonic, having $y$-depencence on the scalar fields, then various mass mixings between the different KK modes are available.  This is useful to give a realistic model.  In our model the candidates of dark matters and their masses in the simplest case without mixings are as follows:
\bea
\hspace{-10mm}
&&
\mbox {HDM:}~~ \nu_{L} = \psi^{(0)}_{L} \,,  
\quad 
m_\mathrm{HDM}=0 \,, \\
\hspace{-10mm}
&&
\mbox {WDM:}~~ N_{R} = \psi^{(0)}_{R} \,, 
\quad 
m_\mathrm{WDM}=\Phi^{(0)}=O(1~\mbox{keV}) \,, \\
\hspace{-10mm}
&&
\mbox {CDM:}~~ \nu^{(1)} = (\psi^{(1)}_{1R} \,, \psi^{(1)}_{2L})~\mbox{and}~(\psi^{(1)}_{2R} \,, \psi^{(1)}_{1L}) \,, 
\quad
m_\mathrm{CDM}=\frac{1}{R}=O(1~\mbox{TeV}) \,, 
\eea
where $\Phi^{(0)}$ is assumed to couple not to $\nu_L$ but to $N_R$.

The reason why the mass of CDM would be $O(\mathrm{TeV})$ is that 
the computer simulations in astronomy and astrophysics predict such a value (for instance, see Ref.~\cite{IAU-DM}). 
The speed of CDM can be determined by mass. It is considered that 
the speed of CDM would be about equal to that of the earth moving around the sun ($230$--$240$km/h). On the other hand, the mass scale of WDM would be 
a middle between the masses of HDM and CDM, and hence we take the mass of 
WDM as $O(1~\mbox{keV})$. It is also known that in computer simulations 
of the galaxy formation, the existence of both WDM and CDM is necessary 
for the time scale to produce them to be close to the observations.  

Moreover, the reasons why we regard the first KK mode of neutrino as 
CDM are that the mass of higher KK modes of neutrino is much larger than 
TeV scale~\cite{Servant and Tait}. Indeed, in the computer simulations, if such a neutrino with its heavy mass is introduced, galaxies cannot be formed, and 
in LHC experiment at CERN the detectable mass is only the lightest one, i.e., 
the first KK mode of neutrino. 
In addition, the higher KK modes of neutrino have the electromagnetic interaction and therefore they cannot be dark matter because dark matter should be neutral. 

The number of the dark matters remaining at present (or the relic density) can be estimated in comparison of the production rate with the expansion rate of the universe.  To make a realistic model various mixings and the higher KK modes should be included.  We here summarize the known results in the simple cases.
The relic density of HDM (neutrino) is known as
\beq
\Omega_\mathrm{HDM} h^2=\frac{\sum m_{\nu}}{93~\mbox{eV}} \,, 
\eeq 
and if WDM was once in a thermal equilibrium, then 
\beq
\Omega_\mathrm{WDM} h^2=\left(\frac{m_{N_{R}}}{93~\mbox{eV}}\right)\left( \frac{T_{N_{R}}}{T_{\nu}} \right)^3 \,,
\eeq 
where $m_{N_{R}}$ and $T_{N_{R}}$ are the mass and temperature 
of the right-handed sterile neutrino, 
and $T_{\nu}=10.75$~K is the decoupling temperatures from the thermal equilibrium.  The relic density of the KK neutrino was estimated in 
Refs.~\cite{Servant and Tait, Kong and Machev}. 
The Figure 2 of \cite{Servant and Tait} shows, if $\Omega_\mathrm{CDM} h^2=0.110 \pm 0.006$, then 
\bea
&&
m_\mathrm{CDM} \approx  0.8-0.9~\mbox{TeV} 
\quad 
\mbox{(for three flavors)} \,, \\
&&
m_\mathrm{CDM} \approx 1.2-1.3~\mbox{TeV} 
\quad 
\mbox{(for one flavor)} \,, 
\eea 
which should be examined including the KK excitations and mixings.

Consequently, the relic density of dark matter consistent with 
observations can be explained 
because in our scenario, the mass of CDM could be 
$O$(TeV), even though dark matters would be a kind of neutrinos 
(namely, HDM is the left-handed neutrino, WDM is the right-handed sterile neutrino, and CDM is the first KK mode of neutrino), 
the mass of which is, in general, considered to be too light 
for them to be candidates for dark matter particles.

\subsection{Direct detectability}

We have to also examine the detectability of dark matters directly 
or indirectly. In Ref.~\cite{Servant:2002hb}, the investigations on the direct detection for the dark matter of the KK particle have been executed. 
Furthermore, the issue on the detectability of dark matters 
has been studied by regarding CDM as the first KK excitation of the neutrino 
in Ref.~\cite{Nishio}. 
How to detect this CDM as well as how to see the seasonal effect through LHC~\cite{LHC}, SK~\cite{Young:1995iv}, and IceCube~\cite{IceCube} have been investigated. 
It has been found that 
the detection of CDM and its seasonal effect would be very difficult using LHC, SK, and IceCube. 
We also note that there has existed the attempt for the detection 
of solar axion~\cite{Solar-Axion}. 

In the following, we explain the investigations on the direct detectability 
of dark matters~\cite{Nishio}. 
Taking the Navarro-Frenk-White (NFW) distribution \cite{NFW} for CDM in the Galaxy, 
\bea
{\rho}_\mathrm{DM}(r)&=&\frac{\delta_{c}\rho_\mathrm{c}}{\left(r/r_\mathrm{s}\right)^{\alpha}\left(1+r/r_\mathrm{s}\right)^{3-\alpha}} \,, 
\quad 
\alpha=1\,,
\eea 
where 
\bea
\delta_{c}&=&\frac{\bigtriangleup_\mathrm{vir}\Omega_{0}}{3}\frac{c_\mathrm{vir}^{3}}{\ln(1+c_\mathrm{vir})-\left[c_\mathrm{vir}/\left(1+c_\mathrm{vir}\right)\right]}\,, \\
\rho_\mathrm{crit}&\equiv&\frac{3{H_{0}}^{2}}{8\pi G_\mathrm{N}}\simeq1.8\times10^{11}h^{2} \frac{M_{\odot}}{\mathrm{Mpc}^{3}} \,, \\
c_\mathrm{vir}&\equiv& \left. \frac{r_\mathrm{vir}(M_\mathrm{halo})}{r_\mathrm{s}(M_\mathrm{halo})}\right|_{z=z_\mathrm{vir}}\,, \\
r_\mathrm{vir}&\equiv& \left(\frac{3M_\mathrm{halo}}{4\pi{\bigtriangleup}_\mathrm{vir}{\Omega}_{0}\rho_\mathrm{crit}}\right)^{\frac{1}{3}}
\approx\frac{1.69}{1+z_\mathrm{vir}}\left(\frac{\bigtriangleup_\mathrm{vir}{\Omega}_{0}}{18\pi^{2}}\right)^{-\frac{1}{3}}\left(\frac{M_\mathrm{halo}}{10^{15}h^{-1}M_{\odot}}\right)^{\frac{1}{3}}\,, \\
\bigtriangleup_\mathrm{vir}&=&18\pi^{2}\Omega(z_\mathrm{vir})^{-1.6}\,.
\eea 
Here, $\delta_\mathrm{c} \approx 0.168$ is the characteristic halo density relative to the critical density $\rho_\mathrm{c} = 9.0738 \times 10^{10} M_{\odot}/\mathrm{Mpc}^3$, the scaling radius $r_\mathrm{s}$ is chosen to $10$ light-years, namely, the radius of the Galactic disc, ${\Omega}_{0}$ is the total density parameter (for the flat space-time, ${\Omega}_{0}=1$), $H_0$ is the current Hubble parameter, $M_{\odot}$ is the solar mass, $r_\mathrm{vir} (M_\mathrm{halo})$ is the virial radius, and $z_\mathrm{vir}$ is the redshift $z$ at the time when the system including the halo mass $M_\mathrm{halo}$ within the virial radius $r_\mathrm{vir} (M_\mathrm{halo})$ virializes. 
The reason why we take a model parameter $\alpha$ as unity is that 
for $\alpha=1$, the large-scale structures can successfully be generated by the computer simulation. 
Since the earth revolves around the sun in the Galaxy, the position of the 
earth from the center of the Galaxy in summer is different from that in 
winter. The maximum of the difference between the position of the earth from 
the center of the Galaxy in the summer (at the summer solstice) and that in the winter (at the winter solstice) is $2$AU.

With the NFW profile for dark matter, the number density of 
the lightest KK particle (LKP) associated with the lightest neutrino 
$\nu^{(1)}$ in the summer and winter read 
\bea
n_{\nu^{(1)}} (\mathrm{summer}) &=& 
4.4 \times 10^{-3} (1- 7.5 \times 10^{-10})/\mathrm{m}^3 \,, 
\label{eq:(30)} \\
n_{\nu^{(1)}} (\mathrm{winter}) &=& 
4.4 \times 10^{-3} (1+ 7.5 \times 10^{-10})/\mathrm{m}^3 \,.
\label{eq:(31)}
\eea
As a consequence, the value of difference of the number density of $\nu^{(1)}$ in the summer from that in the winter is $O(10^{-9})$, and hence 
the detection of this difference is quite hard. 

The cross section for the scattering between $\nu^{(1)}$ and proton $p$ is 
given by 
\bea
\sigma(\nu^{(1)}p) &\approx& G_\mathrm{F}^2 s \,, \\
G_\mathrm{F} &=& 1.2 \times 10^{-5} \mathrm{GeV}^{-2} \,, 
\eea
with $G_\mathrm{F}$ the Fermi constant. 
The square of the reaction energy in the center-of-mass system is described as 
\begin{equation}
s=2E_{q} m_{\nu^{(1)}} +m^2_{\nu^{(1)}}\,, 
\end{equation}
where $E_{q}$ is the energy of a quark $q$ in a proton $p$ which 
reacts a dark matter $\nu^{(1)}$. 
For $s=\left( 5.7 \mathrm{TeV} \right)^2$, we obtain
\beq
\sigma(\nu^{(1)}p) \approx 2.0 \times 10^{-34}\, \mathrm{m}^{2}  \,. 
\eeq
The reaction rate for the interaction per second between 
the dark matter $\nu^{(1)}$ and a proton running with the speed of light $c$ 
 becomes 
\beq
\gamma = \sigma(\nu^{(1)}p) c n_{\nu^{(1)}} \,, 
\eeq
By using Eqs.~(\ref{eq:(30)}) and (\ref{eq:(31)}), we find 
\bea
\gamma (\mathrm{summer}) &=& 3.0 \times 10^{-28} 
\left(1- 7.5 \times 10^{-10}\right)/\mathrm{s}\,, \\
\gamma (\mathrm{winter}) &=& 3.0 \times 10^{-28} 
\left(1+ 7.5 \times 10^{-10}\right)/\mathrm{s}\,. 
\eea

Thus, the numbers of detection of dark matters per month at LHC, $N_\mathrm{LHC}$, was estimated for summer and winter as follows:
\bea
N_\mathrm{LHC}(\mathrm{summer})&=&3 \times 10^7 \left(1- 7.5 \times 10^{-10}\right)/\mathrm{month} \,, \\
N_\mathrm{LHC}(\mathrm{winter})&=&3 \times 10^7 \left(1+ 7.5 \times 10^{-10}\right)/\mathrm{month} \,,
\eea
where the number of protons in the LHC ring has 
been assumed to $3 \times 10^{14}$. 
These are not enough numbers, and therefore it is difficult 
for dark matter to be detected by LHC. 
A way of solving this problem may be to make a number of heavy elements 
run at LHC. 

On the other hand, provided that the solar system runs the dark matter sea, 
where dark matter remains still, with its velocity $v_\mathrm{solar~system} = 
3.0 \times 10^4$ m/s. For example, with the water whose mass is 
$2000$ ton at SK, the speed of proton is $10^{-4} c$, but the number of 
proton is $4 \times 10^{5}$. As a result, the reaction rate is 
$4 \times 10^{11}$/s. 
Accordingly, the corresponding numbers for SK was 
\bea
N_\mathrm{SK}(\mathrm{summer})&=&1.2 \times 10^8 \left(1- 7.5 \times 10^{-10}\right)/\mathrm{month} \,, \\
N_\mathrm{SK}(\mathrm{winter})&=&1.2 \times 10^8 \left(1+ 7.5 \times 10^{-10}\right)/\mathrm{month} \,. 
\eea
If the way of detection is improved, we have the difference of the 
detection numbers in the summer from that in the winter, and 
hence the NFW distribution might be verified. 

We have also confirmed that the resultant numbers of detection of dark matters per month at IceCube are too small. 
In Appendix A, the formulae of the cross sections for the detection of dark matter at LHC, SK, and IceCube derived in Ref.~\cite{Nishio} are presented.

\section{Conclusions}

In the present paper, we have proposed a possible unified scenario of three kinds of dark matters, i.e., Hot, Warm, and Cold Dark Matters, 
into one species (field) in the framework of the KK theory for the five-dimensional space-time. 
Particularly, in our scenario, 
it is provided that HDM, WDM, and CDM would be 
the left-handed neutrino, the right-handed sterile neutrino, and 
the first KK mode of neutrino. 
Furthermore, we have explored the detectability of the first KK neutrino mode 
for CDM by the recent detectors, e.g., LHC, SK, and IceCube. 
It has been found that the detectable numbers of dark matter by 
LHC, SK, and IceCube are still too small. Consequently, some developments 
of the way of detection are necessary. 

Since then the detection techniques have been progressed enormously \cite{Yamashita}. Thus, it is meaningful and significant for us to study the problem again, using Xenon detectors (XEON100~\cite{XENON100}, XMASS~\cite{XMASS} and {\it etc.}) and DAMA/LIBRA~\cite{DAMA-LIBRA}, as wall as the satellite detectors such as PAMELA~\cite{PAMELA}, Planck~\cite{Ade:2013lta, Ade:2013uln} and others, with the necessary information on the nuclear detectors \cite{Nuclear Physics and Detectors} and how cosmic rays moves in the Galaxy \cite{Cosmic rays}.

%
%
\section*{Acknowledgements}
The authors express their thanks to all the members of the particle physics group at Ochanomizu University for their understanding and supporting 
of this work. 
The work has been supported in part by 
the JSPS Grant-in-Aid for Young Scientists (B) \# 25800136 (K.B.).

\section*{Appendix}

In this Appendix, we write 
the formulae of the cross sections obtained in Ref.~\cite{Nishio} in the study of dark matter detection at LHC, SK, and IceCube are still useful. 

We denote LKP 
and LKP associated with the lepton ($\ell$) for $\nu^{(1)}$ 
by 
$\ell^{(1)}$. 
We need the cross section for the reaction between $\nu^{(1)}$ and a proton 
so that we can detect $\nu^{(1)}$ at LHC because at LHC the proton beam is 
generated. It is also necessary to know the cross section for the reaction between $\nu^{(1)}$ and a quark ($u$ or $d$) because a proton consists of two $u$ quarks and one $d$ quark such as $(uud)$. There exist two reactions: One is via a 
$W$ boson $(1,\,2)$, and the other is through a $Z$ boson $(3,\,4)$. 
The former is called Charged Current (CC) process, whereas the latter is called Neutral Current (NC) process. 
We summarize the reactions in the following: 

(1) $\nu^{(1)}+d \rightarrow \ell^{(1)}+u$
\beq
\sigma_\mathrm{CC}({\nu^{(1)}}+d\rightarrow{\ell^{(1)}}+u)=\frac{g^{4}}{64\pi}\frac{(s-{m^{2}}_{\ell^{(1)}})^{2}}{{m^{2}}_\mathrm{W}}\frac{1}{(s-{m^{2}}_{\nu^{(1)}})(s-{m^{2}}_{\ell^{(1)}})+s{{m_\mathrm{W}}^{2}}} \,.
\eeq

(2) ${\bar\nu^{(1)}}+u \rightarrow {\bar\ell^{(1)}}+d$
\bea
&&
\sigma_\mathrm{CC}({\bar\nu^{(1)}}+u \rightarrow {\bar\ell^{(1)}}+d)=\frac{g^{4}}{64\pi}\frac{1}{(s-{m^{2}}_{\nu^{(1)}})^{2}} 
\nonumber \\
&&
{}\times \left[\frac{(s-{m^{2}}_{\nu^{(1)}})(s-{m^{2}}_{\ell^{(1)}})(s+{{m^{2}}_\mathrm{W}}-{{m^{2}}_{\nu^{(1)}}})(s+{{m^{2}}_\mathrm{W}}-{{m^{2}}_{\ell^{(1)}}})}{{m^{2}}_\mathrm{W}\{(s-{m^{2}}_{\nu^{(1)}})(s-{m^{2}}_{\ell^{(1)}})+s{m^{2}}_\mathrm{W}\}}
\right. 
\nonumber \\
&& \left.
{}+(2s+2{m^{2}}_\mathrm{W}-{m^{2}}_{\ell^{(1)}}-{m^{2}}_{\nu^{(1)}})
\ln \left|1+\frac{(s-{m^{2}}_{\nu^{(1)}})(s-{m^{2}}_{\ell^{(1)}})}{s{m^{2}}_\mathrm{W}}\right| 
\right. 
\nonumber \\
&& \left.
{}+\frac{1}{s}(s-{m^{2}}_{\nu^{(1)}})(s-{m^{2}}_{\ell^{(1)}})\right] \,.
\eea

(3) $\nu^{(1)}+q \rightarrow \nu^{(1)}+q$
\bea
&&
\sigma_\mathrm{NC}({\nu^{(1)}}+q \rightarrow {\ell^{(1)}}+q)=\frac{1}{64\pi}\left(\frac{g}{{\cos\theta}_\mathrm{W}}\right)^{4}\frac{1}{(s-{m^{2}}_{\nu^{(1)}})^{2}}
\nonumber \\
&&
{}\times \Biggl[\left\{\left(\frac{g_{V}+g_{A}}{2}\right)^{2}(s-{m^{2}}_{\nu^{(1)}})^{2}+\left(\frac{g_{V}-g_{A}}{2}\right)^{2}{(s+{m^{2}}_{Z}-{m^{2}}_{\nu^{(1)}})}^{2}\right\}
\nonumber \\
&&
{}\times\frac{(s-{m^{2}}_{\nu^{(1)}})^{2}}{{m^{2}}_\mathrm{W}\{(s-{m^{2}}_{\nu^{(1)}})^{2}+s{m^{2}}_{Z}\}}
\nonumber \\
&&
{}+\left\{\left(\frac{g_{V}+g_{A}}{2}\right)^{2}\times2(s+{m^{2}}_{Z}-{m^{2}}_{\nu^{(1)}})\right\}
\ln \left| 
1+\frac{(s-{m^{2}}_{\nu^{(1)}})^{2}}{s{m^{2}}_{Z}} \right|
\nonumber \\
&&
{}+\left(\frac{g_{V}+g_{A}}{2}\right)^{2}\times\frac{(s-{m^{2}}_{\nu^{(1)}})^{2}}{s} \Biggr] \,.
\eea

(4) $\bar\nu^{(1)}+q \rightarrow \bar\nu^{(1)}+q$ 

This cross section is  obtained from $\sigma_\mathrm{NC}(\nu^{(1)}+q \rightarrow  \ell^{(1)} +q)$ by the replacement $(g_{V}+g_{A}) \leftrightarrow (g_{V}-g_{A})$.

Here, we have 
\begin{eqnarray}
\frac{g_{A}+g_{V}}{2}&=&\left\{ \begin{array}{ll}
\frac{1}{2}-\frac{2}{3}\sin^2 \theta_\mathrm{W} \quad \mbox{for} \,\, u \,\, \mbox{quark} \,, \\
-\frac{1}{2}+\frac{1}{3} \sin^2 \theta_\mathrm{W} \quad \mbox{for} \,\, d \,\, \mbox{quark} \,,
\end{array} 
\right. \\
\frac{g_{A}-g_{V}}{2}&=&\left\{ \begin{array}{ll}
-\frac{2}{3}\sin^2 \theta_\mathrm{W} \quad \mbox{for} \,\, u \,\, \mbox{quark} 
\,, \\
\frac{1}{3}\sin^2 \theta_\mathrm{W} \quad \mbox{for} \,\, d \,\, \mbox{quark} 
\,, 
\end{array} 
\right.
\end{eqnarray} 
where 
$\sin^2 \theta_\mathrm{W} \approx 0.23$ with $\theta_\mathrm{W}$ 
the Weinberg angle, $g^{2} \approx 0.43$ with $g$ the coupling constant, 
$m_\mathrm{W} \approx 80.4$ GeV is the W boson mass, and $m_{Z} \approx 91.2$ GeV is the Z boson mass. 
At the LHC experiment, if the energy of proton $p$ is $7$ TeV, 
the one-third of this energy, i.e., the averaged value by three 
quarks in the proton is presented to each quark. 
Accordingly, we acquire 
\begin{equation}
E_{q} \approx \frac{7}{3} \, \mbox{TeV} \,. 
\end{equation}
For $m_{\nu^{(1)}} = 1$TeV, we find 
\begin{equation}
s=5.7 \, \mbox{TeV}^2 \,. 
\end{equation} 
%

%
%

\end{document}